# Missing Data: A Comparison of Neural Network and Expectation Maximisation Techniques

Fulufhelo V. Nelwamondo, Shakir Mohamed and Tshilidzi Marwala

[*]School of Electrical and Information Engineering, University of the Witwatersrand
Private Bag 3, Wits, 2050, South Africa,
e-mail: f.nelwamondo@ee.wits.ac.za, s.mohamed@ee.wits.ac.za  t.marwala@ee.wits.ac.za

**Abstract**

Two techniques have emerged from the recent literature as candidate solutions to the problem of missing data imputation, and these are the Expectation Maximisation (EM) Algorithm and the auto-associative Neural Networks and Genetic Algorithms combination. Both these techniques have been discussed individually and their merits discussed at length in the available literature, but up to this point, they have not been compared with each other. This paper provides this comparison, using data sets of an industrial power plant, an industrial winding process and HIV sero-prevalence survey data. Results show that Expectation Maximization is suitable and performs better in cases where there is little or no interdependency between the input variables, whereas the auto-associative neural network and genetic algorithm combination is suitable when there is some inherent non-linear relationships between some of the given variables.

# 1. INTRODUCTION

Databases such as those which store measurement or medical data may become subject to missing values either in the data acquisition or data-storage process. Problems in a sensor, a break in the data transmission line or non-response to questions posed in a questionnaire are prime examples of how data can go missing. The problem of missing data poses a difficulty to the analysis and decision making processes which depend on this data, requiring methods of estimation which are accurate and efficient. Various techniques exist as a solution to this problem, ranging from data deletion to methods employing statistical and artificial intelligence techniques to impute for missing variables. However, some statistical methods, like mean substitution have a high likelihood of producing biased estimates[1] or make assumptions about the data that may not be true, affecting the quality of decisions made based on this data.

The estimation of missing input vector elements in real time processing applications requires a system that possesses the knowledge of certain characteristics such as correlations between variables, which are inherent in the input space. Computational intelligence techniques and maximum likelihood techniques do possess such characteristics and as a result are important for imputation of missing data. This paper compares two approaches to the problem of missing data estimation. The first technique is based on the current state of the art approach to this problem, that being the use of Maximum Likelihood (ML) and Expectation Maximisation (EM)[2]. The second approach is the use of a system based on auto-associative neural networks and the Genetic Algorithm as discussed by Adbella and Marwala[3]. The estimation ability of both of these techniques is compared, based on three datasets and conclusions are made.

## 2. BACKGROUND

### *2.1 Missing Data*

Real time processing applications that are highly dependent on the data often suffer from the problem of missing input variables. Various heuristics of missing data imputation such as mean substitution and hot deck imputation also depend on the knowledge of how data points become missing. There are several reasons why the data may be missing, and as a result, missing data may follow an observable pattern. Exploring the pattern is important and may lead to the possibility of identifying cases and variables that affect the missing data[4]. Having identified the variables that predict the pattern, a proper estimation method can be selected.

According to Little and Rubin[5] and Burk[6], there are three types of missing data mechanisms. These mechanisms can be distinguished by considering figure 1 which shows a data pattern with variables $X = \{X_1, X_2, \ldots X_p\}$ and $Y$ which has some missing data. Variables $X$ and $Y$ represent columns of a table in the database, with the column $Y$ being the column with missing data. Considering $X$ and $Y$ as random variables, the three categories of missing data are:

**Missing Completely at Random (MCAR)** which occurs if the missing value for the input vector has no dependence on any other variable in the database such that the inputs with missing entries are the same as the complete inputs. That is, the probability of data $Y$, being missing is not dependant on either $X$ or $Y$, i.e. is not dependant on either missing or complete values in the same record or any other record in the database.

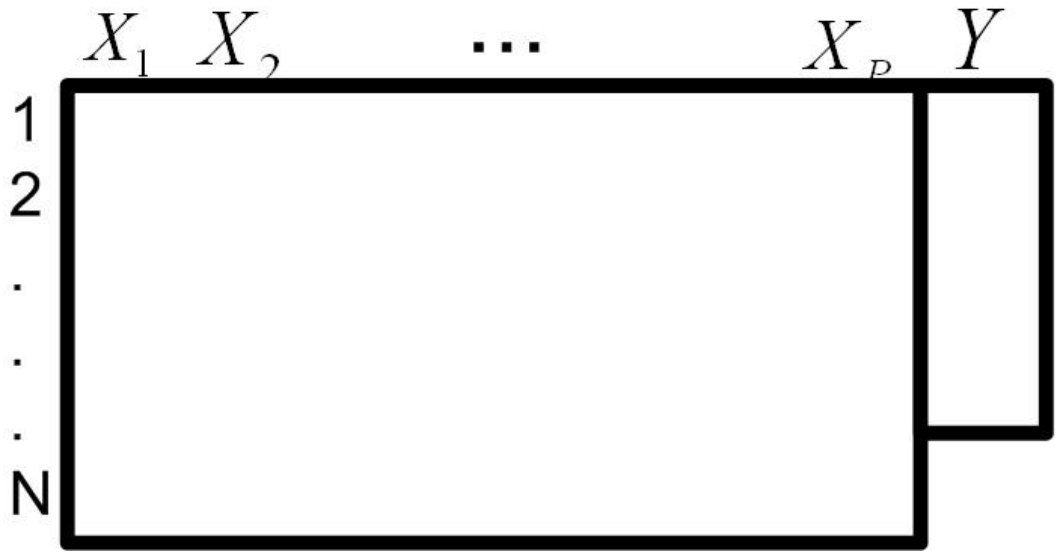

Figure 1. Pattern of missing data in a rectangular data set. Rows correspond to records in the database and the columns correspond to variables or fields of the data set [2].

**Missing at Random (MAR)** occurs if the missing value for the input vector has dependence on other variables in the data set, such that the pattern in which the data becomes missing is traceable. That is, the probability of data *Y* being missing is dependant only on *X* the existing values in the database and not on any missing data.

**Missing Not at Random (MNAR)** occurs when the missing value for the input vector depends on the other missing values such that the existing data in the database cannot be used to approximate the missing values. This is also known as the non-ignorable case. The probability that *Y* is missing is dependant on the missing data.

This work assumes that data is missing at random. This implies that we expect the missing values to be deducible in some complex manner from the remaining data.

*2.2. Autoencoder Neural Networks*

Autoencoders, also known as auto-associative neural networks, are neural networks trained to recall the input space. Thompson *et al*[7] distinguish two primary features of an autoencoder network, namely the auto-associative nature of the network and the presence of a bottleneck that occurs in the hidden layers of the network, resulting into a butterfly-like structure. In cases where it is necessary to recall the input, autoencoders are preferred due to their remarkable ability to learn certain linear and non-linear interrelationships such as correlation and covariance inherent in the input space. Autoencoders project the input onto some smaller set by *intensively squashing* it into smaller details. The optimal number of the hidden nodes of the autoencoder, though dependent on the type of application, must be smaller than that of the input layer[7]. Autoencoders have been used in various applications including the treatment of missing data problem by a number of researchers[3, 8-10].

In this paper, auto-encoders are constructed using the Multi-layer perceptrons (MLP) networks and trained using back-propagation. MLPs are feed-forward neural networks with an architecture comprising of the input layer, the hidden layer and the output layer. Each layer is formed from smaller units known as neurons. Neurons in the input layer receive the input signals $\bar{x}$ and distribute them forward to the network. In the next layers, each neuron receives a signal, which is a weighted sum of the outputs of the nodes in the previous layer. Inside each neuron, an activation function is used to control the input. Such a network determines a non-linear mapping from an input vector to the output vector, parameterised by a set of network weights, which we refer to as the vector of weights $\vec{W}$. The structure of an autoencoder constructed using an MLP network is shown in figure 2.

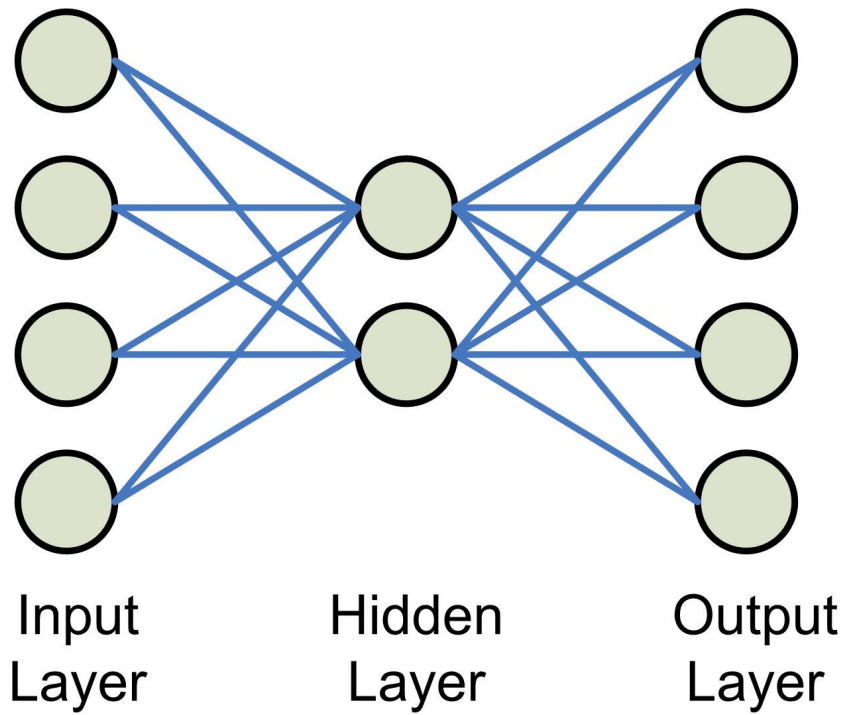

Figure 2. The structure of a four-input four-output autoencoder

The first step in approximating the weight parameters of the model is finding the approximate architecture of the MLP, where the architecture is characterized by the number of hidden units, the type of activation function, as well as the number of input and output variables. The second step estimates the weight parameters using the training set[11]. Training estimates the weight vector $\vec{W}$ to ensure that the output is as close to the target vector as possible. The problem of identifying the weights in the hidden layers is solved by maximizing the probability of the weight parameter using Bayes' rule[7] as follows:

$$P(\vec{W}|D) = \frac{P(D|\vec{W})P(\vec{W})}{P(D)} \quad (1)$$

where, D is the training data, $P(D|\vec{W})$ is called the evidence term that balances between fitting the data well and helps in avoiding overly complex models whereas $P(\vec{W})$ is the prior probability of $\vec{W}$. The input is transformed from *x* to the middle layer, *a*, using weights $W_{ij}$ and biases $b_i$ as follows[7]:

$$a_j = \sum_{i=1}^{d} W_{ji} x_i + b_j \quad (2)$$

where *j=1* and *j=2* represent the first and second layer respectively. The input is further transformed using the activation function such as the hyperbolic tangent (*tanh*) or the sigmoid in the hidden layer. More information on neural networks can be found in the book by Bishop[12].

*2.3 Genetic Algorithms*

There are various optimization techniques such as fast simulated annealing, ant colony optimization, genetic algorithms and particle swam optimisation that are all aimed at optimizing some variables to adhere to some target function. Some of these methods converge at local optimal solutions than the required global optimal solutions. Although stochastic in nature, GA converges at a global optimal solution. GA's use the concept of
Survival of the fittest over consecutive generations to solve optimization problems[13] . As in biological evolution, the fitness of each population member in a generation is evaluated to determine whether it will be used in the breeding of the next generation. In creating the next generation, the use of techniques (such as inheritance, mutation, natural selection, and recombination) common in the field of evolutionary biology are employed. The GA algorithm implemented in this paper uses a population of string chromosomes, which represent a point in the search space[13]. In this paper, all these parameters were empirically determined using exhaustive search methods. GA is implemented by following three main procedures which are selection, crossover and mutation.

## 3. NEURAL NETWORKS AND GENETIC ALGORITHM FOR MISSING DATA

The method used here combines the use of auto-associative neural networks with genetic algorithms to approximate missing data. This method has been used to approximate missing data in a database by Abdella and Marwala[3]. A genetic algorithm is here used to *estimate* the missing values by optimizing an objective function as presented shortly in this section. The complete vector combining the guessed and the observed values is fed into the autoencoder as input and as shown in figure 3. Symbols $X_k$ and $X_u$ represent the known variables and the unknown or missing variable respectively. The combination of $X_k$ and $X_u$ represent the full input space

Considering that the method uses an autoencoder, one will expect the input to be very similar to the output for a well chosen architecture of the autoencoder. This is, however, only expected on a data set similar to the problem space from which the inter-correlations have been captured. The difference between the target and the actual output is used as the error and this error is defined as follows:

$$\varepsilon = \vec{x} - f(\vec{W}, \vec{x}) \tag{3}$$

where $\vec{x}$ and $\vec{W}$ are input and weight vectors respectively. To make sure the error function is always positive, the square of the equation is used. This leads to the following equation:

$$\varepsilon = (\vec{x} - f(\vec{W}, \vec{x}))^2 \tag{4}$$

Since the input vector consist of both the known, $X_k$ and unknown, $X_u$ entries, the error function can be written as follows:

$$\varepsilon = \left( \begin{Bmatrix} X_k \\ X_u \end{Bmatrix} - f\left( \begin{Bmatrix} X_k \\ X_u \end{Bmatrix}, w \right) \right)^2 \tag{5}$$

and this equation is used as the objective function that is minimized using GA.

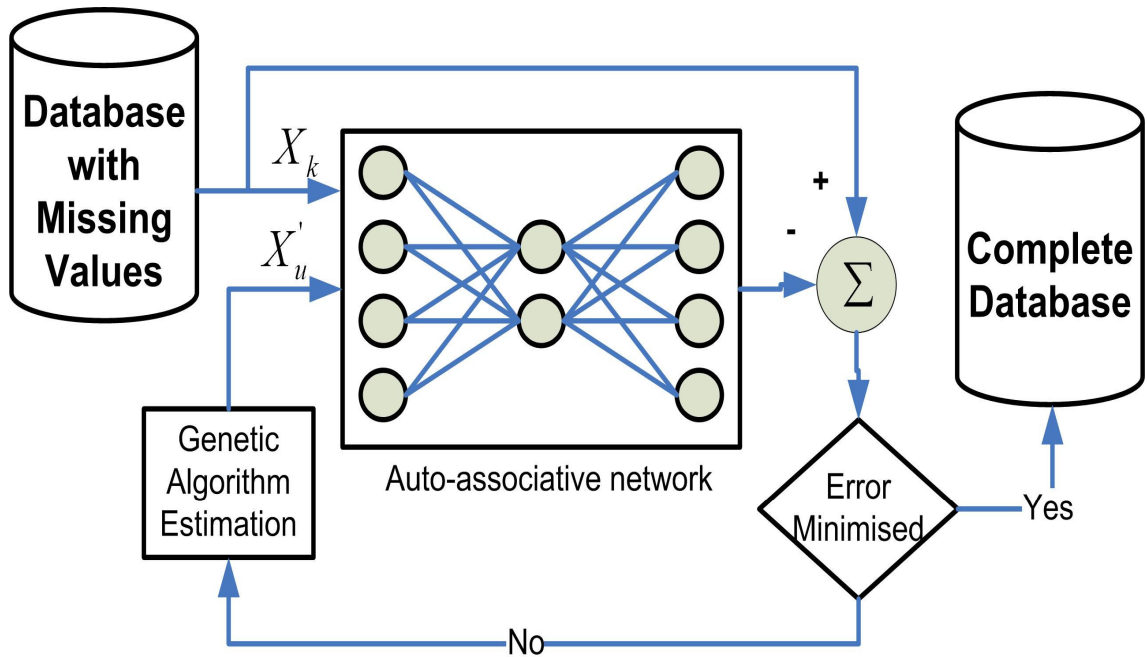

Figure 3. Autoencoder and GA Based missing data estimator structure

## 4. MAXIMUM LIKELIHOOD

The maximum likelihood approach to approximating missing data is a very popular technique[14-15] and is based on a precise statistical model of the data. The model most commonly used is the multivariate, Gaussian mixture model while the maximum likelihood method is applied for the task of imputing the missing values. Likelihood methods may be categorized into 'Single

Imputations' and 'Multiple Imputations'[4, 14-15]. This paper will only consider Single Imputations and as a result, the EM algorithm will be used and is discussed further below.

*4.1. Expectation Maximization for Missing Data*

The expectation-maximization algorithm was originally introduced by Dempster *et al*[16] in 1977 and was aimed at overcoming problems associated with Maximum Likelihood methods. Expectation Maximisation combines statistical methodology with algorithmic implementation and has gained much attention recently in various missing data problems. Expectation Maximization has also been proven to work better than methods such as listwise, pairwise data deletion, and mean substitution because it assumes incomplete cases have data missing at random rather than missing completely at random[17-18].

The EM algorithm is a general technique for fitting models to incomplete data. Expectation Maximization capitalizes on the relationship between missing data and the unknown parameters of a data model. If we knew the missing values, then estimating the model parameters would be straightforward. Similarly, if we knew the parameters of the data model, then it would be possible to obtain unbiased predictions for the missing values. This interdependence between model parameters and missing values suggests an iterative method where we first predict the missing values based on assumed values for the parameters, use these predictions to update the parameter estimates, and repeat. The sequence of parameters converges to maximum-likelihood estimates that implicitly average over the distribution of the missing values.

## 5. EXPERIMENTAL EVALUATION

*5.1 Data Analysis*

The Expectation Maximization and the neural network coupled with genetic algorithms approaches for approximating missing data were compared in three different data sets. The data sets used are briefly described below:

**A) Power plant data**

The first data set used is the data of a 120 MW power plant in France[19], under normal operating conditions. This data set comprises five inputs, namely: gas flow, turbine valves opening, super heater spray flow, gas dampers and air flow. Sampling of the data was done every 1228.8 seconds and a total of 200 instances were recorded. An extract of the data without any missing values is shown in Table 1.

Table 1. Set of power plant measurements under normal operating conditions

| Gas Flow | Turbine | Heater | Gas Dampers | Air Flow |
|---|---|---|---|---|
| 0.11846 | 0.089431 | 0.11387 | 0.6261 | 0.076995 |
| 0.10859 | 0.082462 | 0.11284 | 0.6261 | 0.015023 |
| 0.099704 | 0.19919 | 0.14079 | 0.62232 | 0.061972 |
| 0.092794 | 0.19164 | 0.12733 | 0.6261 | 0.059155 |
| 0.0888845 | 0.30023 | 0.13768 | 0.6261 | 0.028169 |
| 0.0875858 | 0.63182 | 0.074834 | 0.63052 | 0.079812 |

The data was split into training and testing data sets. Due to the limited data available, one seventh of the data was kept as the test set, with the remaining being considered for training. For easy comparison with the NN-GA, the training and testing data for the EM were combined into a single file, with the testing data appearing at the end of the file. This separation ensured that both the EM and the neural network and genetic algorithm approach NN-GA) testing were compared using the same amount of testing data and that their respective models are built from the same amount of ``training'' data. The data was transformed using a min-max normalization to [0,1] before use, to ensure that the data is within the active range of the activation function of the neural network

**B) HIV database**

The data used in this test was obtained from the South African antenatal sero-prevalence survey of 2001. The data for this survey is obtained from questionnaires answered by pregnant women visiting selected public clinics in South Africa. Only women participating for the first time in the survey were eligible to answer the questionnaire.

Data attributes used in this study are the HIV status, Education level, Gravidity, Parity, Age Group and Age Gap. The HIV status is represented in a binary form, where 0 and 1 represent negative and positive respectively. The education level was measured using integers representing the highest grade successfully completed, with 13 representing tertiary education. Gravidity is the number of pregnancies, complete or incomplete, experienced by a female, and this variable is represented by an integer between 0 and 11. Parity is the number of times the individual has given birth and multiple births are considered as one birth event. Both parity and gravidity are important, as they show the reproductive activity as well as the reproductive health state of the woman. Age gap is a measure of the age difference between the pregnant woman and the prospective father of the child. A sample of this data set is shown in Table 2. The data consists of 5776 instances and this data was divided into two subsets, namely, training and testing data sets as training was done in the Bayesian framework. Testing was done with 776 instances.

Table 2. Extract of the HIV database used, without missing values

| HIV | EDUC | GRAVIDITY | PARITY | AGE | AGE GAP |
|-----|------|-----------|--------|-----|---------|
| 0   | 7    | 10        | 9      | 35  | 5       |
| 1   | 10   | 2         | 1      | 20  | 2       |
| 1   | 10   | 6         | 5      | 40  | 6       |
| 0   | 5    | 4         | 3      | 25  | 3       |

**C) Data from an industrial winding process**

The third data set used here represents a test setup of an industrial winding process and the data can be found in[19]. The main part of the plant is composed of a plastic web that is unwound from the first reel (unwinding reel), goes over the traction reel and is finally rewound on the rewinding reel as shown in figure 4.

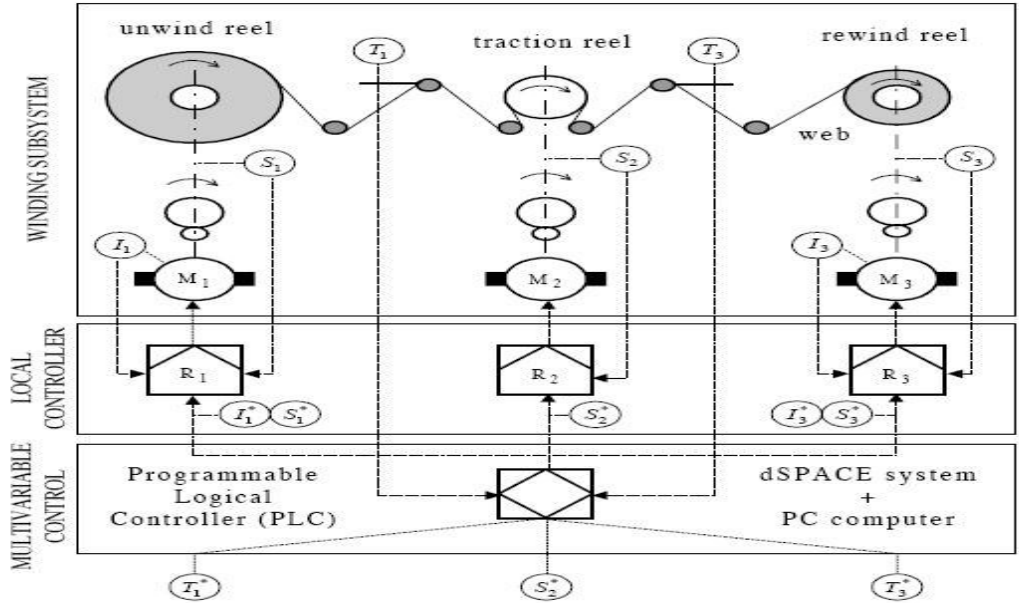

Figure 4.  The graphic representation of the winding plot system [20]

As shown in the figure, reels 1 and 3 are coupled with a DC-motor that is controlled with input set-point currents I1 and I3. The angular speed of each reel (S1, S2 and S3) and the tensions in the web between reel 1 and 2 (T1) and between reel 2 and 3 (T3) are measured by dynamo tachometers and tension meters. The full data set has 2500 instances, sampled every 0.1 seconds. In this study, testing was done with 500 instances while the training set and the validation set for the neural network consisted of 1500 and 500 instances respectively.  The inputs to the winding system are the angular speed of reel 1 (S1), reel 2 (S2), reel 3 (S3), the set point current at motor 1 (I1) and at motor 2 (I3)} as shown in Figure 4. A more detailed description of the data can be found in[20].

*5.2 Performance Analysis*

The effectiveness of the missing data system is evaluated using the correlation coefficient and the relative prediction accuracy. The correlation coefficient will be used as a measure of similarity between the prediction and the actual data. The correlation coefficient, $r$ is computed as:

$$r = \frac{\sum_{i=1}^{n}(x_i - \bar{x}_i)(\hat{x}_i - \bar{\hat{x}}_i)}{[\sum_{i=1}^{n}(x_i - \bar{x}_i)^2 \sum_{i=1}^{n}(x_i - \bar{x}_i)(\hat{x}_i - \bar{\hat{x}}_i)^2]^{1/2}} \quad (6)$$

where $\hat{x}$ represents the approximated data, $x$ is the actual data while $\bar{x}$ represents the mean of the data. The relative prediction accuracy is defined as:

$$Error = \frac{n_\tau}{N} \times 100\% \quad (7)$$

where $n_\tau$ is the number of predictions within a certain tolerance percentage of the missing value. In this paper, a tolerance of 10% is used. The 10% was arbitrarily chosen with an assumption that it is the maximum acceptable margin for error in the applications considered. This error analysis can be interpreted as a measure of how many of the missing values are predicted within the tolerance and the tolerance can be made to be any value depending on the sensitivity of the application.

## 6. EXPERIMENTAL RESULTS AND DISCUSSION

This section presents the experimental results obtained by using both of the approaches described in Section 3 and Section 4. We evaluate predictability within 10% of the target value. The evaluation is computed by determining how much of the test sample was estimated within the given tolerance. We first present the results of the test done using the power plant data set.

For the experiment with the power plant data, the Neural Network--Genetic Algorithm (NN-GA) system was implemented using an autoencoder network trained with 4 hidden nodes for 200 training epochs. The Genetic Algorithm was implemented using the floating point representation for 30 generations, with 20 chromosomes per generation. The mutation rate was set to a value of 0.1. As mentioned earlier, the GA parameters were empirically determined. The correlation coefficient, and the accuracy within 10% of the actual values are given in table 3.

Table 3. Results of comparative testing using power plant data

| Variable | Correlation | | 10% | |
|---|---|---|---|---|
| | **Corr EM** | **Corr NN-GA** | **EM** | **NN-GA** |
| Gas Flow | - | 0.9790 | - | 21.43 |
| Turbine | 0.7116 | 0.8061 | 14.29 | 14.29 |
| Heater | 0.7218 | 0.6920 | 7.14 | 28.57 |
| Gas dumper | -0.4861 | 0.5093 | 3.57 | 10.71 |
| Air Flow | 0.6384 | 0.8776 | 10.71 | 7.14 |

It can be seen from the results that EM failed to make a prediction for column 1 in this data set. The reason is that for EM to make a prediction, the prediction matrix needs to be positive definite[21]. The major cause of this is when one variable is linearly dependent on another variable. This linear dependency may sometimes exist not between the variables themselves, but between elements of moments such as the mean, variances, covariances and correlations[21]. Other reasons for this cause include errors while reading the data, initial values and many more. This problem can be solved by deleting variables that are linearly dependent on each other or by using Principal Components to replace a set of collinear variables with orthogonal components. Seminal work on dealing with "not positive definite matrices" was done by Wothke[21].

The results show that the NN-GA methods is able to impute missing values with higher accuracy of prediction for most cases and this is shown in the graph of figure 5. The lack of high accuracy predictions for both estimation techniques though suggests some degree of difficulty in estimating the missing variables based in the given set of data.

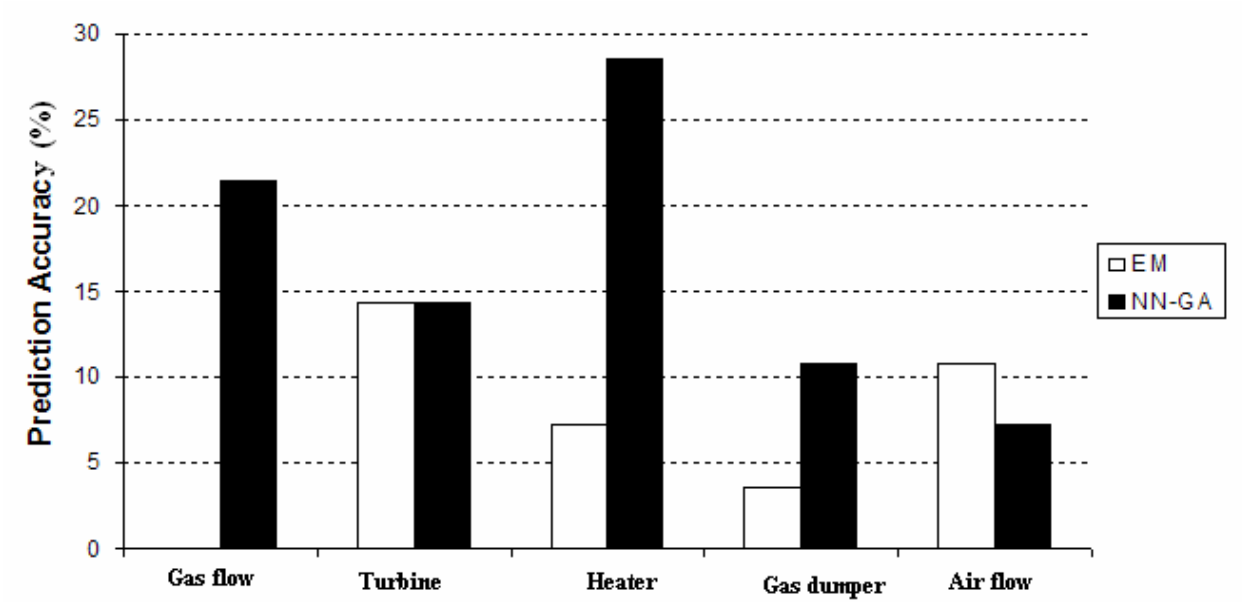

Figure 5. Graphical comparison of estimation accuracy using 10% tolerance using power plant data

For neural networks, it is observable that the quality of estimation in each input variable depends on the existence of some form of correlation between variables in the input space such that this linear or non-linear relationship can be discovered by the neural networks and used to give higher accuracy imputations. The EM algorithm also requires that the data not to be linearly dependent on some variables within the input space, as demonstrated by the need for positive definite matrices. Before commencing with the experiment, data was tested for correlation. This testing involved finding out if any variable in the data is somehow strongly related to any other variable in the data.

The results obtained using the HIV database are presented below. Figure 6 shows the results obtained when predicting missing variables on the HIV data set within 10% tolerance. Results here clearly show that Expectation Maximization performs better than the NN-GA method for the prediction of variables such as Education, Parity, Age and Age gap. Unlike with the power plant database, results here show that EM is better than Neural Network for prediction of variables in the HIV data set in this study. Since this is a social science database, the reason for poor performance of the NN-GA can be either that the variables are not sufficiently representative of the problem space to produce an accurate imputation model, or that people were not very honest in answering the questions in the questionnaire, leading to less dependability of variables on each other.

We lastly show the results obtained from the industrial winding process. EM and NN-GA approaches are compared and the results are shown in Figure 7. Again the results obtained in this section show that for some variables the EM algorithm produces better imputed results, while in others the NN-GA system was able to produce a better imputation result. From the observed data the predicted values are not very correlated to the actual missing variables. The possible explanation to this is that the missing data is not interdependent on itself, but to other variables in the data. Table 4 shows the correlation coefficients. As for the other data sets, the problem of the non-positive definite matrix when imputing values for column 1 prevented the EM algorithm from being used to estimate the missing data.

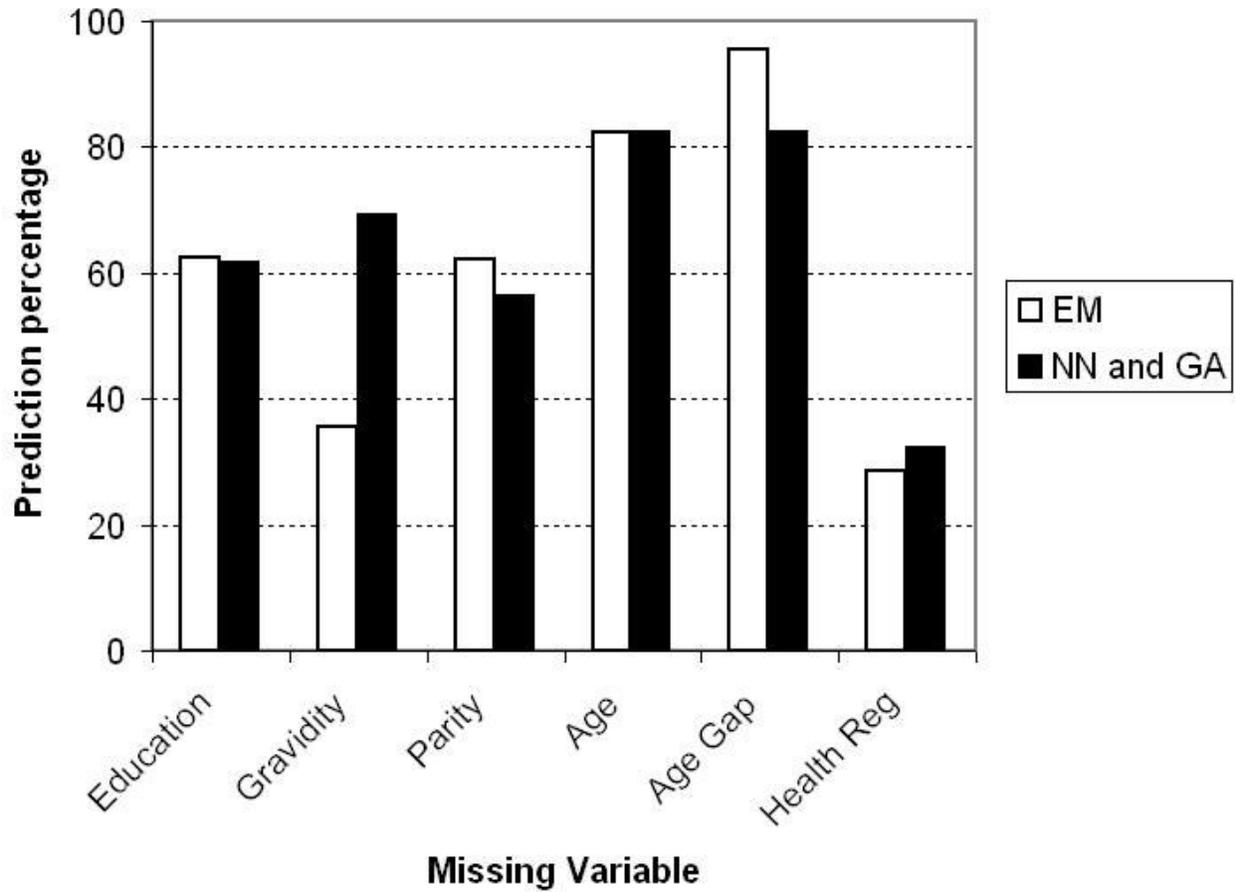

Figure 6. Prediction within 10 % tolerance of missing variable in the HIV data

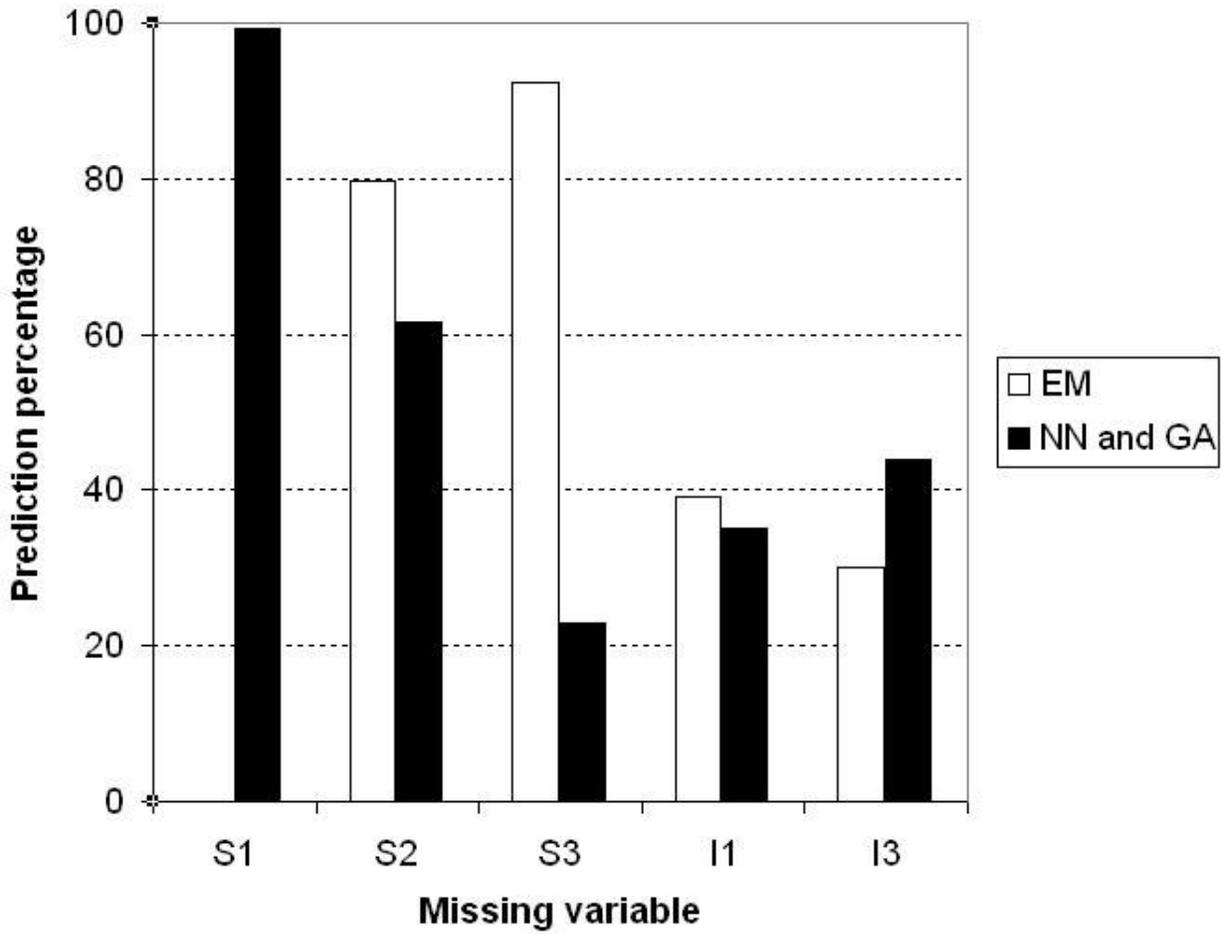

Figure 7. Prediction within 10 % tolerance of missing variable in the industrial winding

Process

Table 4. The correlation coefficients between actual and predicted for the winding process data

|       | S1    | S2     | S3    | I1    | I2      |
|-------|-------|--------|-------|-------|---------|
| NN-GA | 0.203 | 0.229  | 0.159 | 0.038 | 0.117   |
| EM    | -     | -0.003 | 0.009 | -0.05 | -0.0007 |

# 7. CONCLUSIONS

This paper investigated and compared the maximum likelihood approach and neural networks and GA combination approach for missing data approximation. In one approach, an auto-associative neural network was trained to predict its own input space. Genetic algorithms were used to approximate the missing data. On the other hand, expectation maximization was implemented for the same problem. The results show that for some variables the EM algorithm is able to produce a better imputation accuracy, while for the other variables the neural network-genetic algorithm system is better. Thus the imputation ability of one method over another seems highly problem dependant. Findings also showed that EM seems to perform better in cases where there is a very little dependency among the variables, which is contrary to the performance of the neural network approach.